\documentclass[aps,twocolumn,showpacs,preprintnumbers,nofootinbib,prl,superscriptaddress,groupedaddress,10pt]{revtex4-1}
\pdfoutput=1
\makeatletter
\def\l@subsubsection#1#2{}
\def\l@subsubsubsection#1#2{}
\makeatother

\usepackage{graphicx,amssymb,amsmath,amsthm,amsfonts,epsfig,epsf,fixmath}
\usepackage[usenames]{color}
\usepackage{epstopdf}
\usepackage{mathtools}
\usepackage{aas_macros}
\usepackage{bm}
\usepackage{dcolumn}
\usepackage{latexsym}
\usepackage{rotating}
\usepackage{longtable}

\usepackage{enumerate}
\usepackage{tensor,multirow}
\usepackage{url}
\usepackage[linktocpage]{hyperref}

\def\nn{\nonumber}

\newcommand{\pt}[1]{\left(#1\right)}

\newcommand{\pmat}{\begin{pmatrix}}
\newcommand{\fpmat}{\end{pmatrix}}
\newcommand{\eq}{\begin{equation}}
\newcommand{\feq}{\end{equation}}
\newcommand{\cas}{\begin{cases}}
\newcommand{\fcas}{\end{cases}}

\newcommand{\eqarray}{\begin{eqnarray}}
\newcommand{\feqarray}{\end{eqnarray}}

\newcommand{\be}{\beta}

\def\be{\begin{equation}}
\def\ee{\end{equation}}
\def\bea{\begin{eqnarray}}
\def\eea{\end{eqnarray}}

\newcommand{\half}{\frac{1}{2}}

\begin{document}
\title{Distinguishing fuzzballs from black holes through their multipolar structure}
\author{
Massimo Bianchi$^{1}$,
Dario Consoli$^{2}$,
Alfredo Grillo$^{1}$,
Jos\`e Francisco Morales$^{1}$,
Paolo Pani$^{3}$,
Guilherme Raposo$^{3}$
}

\affiliation{$^{1}$ Dipartimento di Fisica,  Università di Roma ``Tor Vergata"  \& Sezione INFN Roma2, Via della ricerca scientifica 1, 
00133, Roma, Italy}
\affiliation{$^{2}$ Mathematical Physics Group, University of Vienna, Boltzmanngasse 5 1090 Vienna, Austria}
\affiliation{$^{3}$ Dipartimento di Fisica, ``Sapienza" Universit\`a di Roma \& Sezione INFN Roma1, Piazzale Aldo Moro 
5, 
00185, Roma, Italy}

\begin{abstract}
Within General Relativity, the unique stationary solution of an isolated black hole is the Kerr spacetime, which has a 
peculiar multipolar structure depending only on its mass and spin. We develop a general method to extract the multipole 
moments of arbitrary stationary spacetimes and apply it to a large family of horizonless microstate geometries. The 
latter can break the axial and equatorial symmetry of the Kerr metric and have a much richer multipolar 
structure, which provides a portal to constrain fuzzball models phenomenologically. We find numerical evidence that
all multipole moments are typically larger (in absolute value) than those of a 
Kerr black hole with the same mass and spin. Current measurements of the quadrupole moment of black-hole 
candidates could place only mild constraints on fuzzballs, while future gravitational-wave detections of extreme 
mass-ratio inspirals with the space mission LISA will improve these bounds by orders of magnitude. 
\end{abstract}

\maketitle 

\noindent{{\bf{\em Introduction.}}}
Owing to the black-hole (BH) uniqueness and no-hair
theorems~\cite{Carter71,Hawking:1973uf} (see also Refs.~\cite{Heusler:1998ua,Chrusciel:2012jk,Robinson}),
within General Relativity~(GR) any stationary BH in isolation is also axisymmetric and its multipole 
moments\footnote{For a generic spacetime the multipole moments of order $\ell$ are rank-$\ell$ tensors, which reduce to 
scalar quantities, ${\cal M}_\ell$ and ${\cal S}_\ell$, in the axisymmetric case. See below for the 
general definition.} satisfy an elegant relation~\cite{Hansen:1974zz}, 
\begin{equation}
 \mathcal{M}_\ell^{\rm BH}+{\rm i }  \mathcal{S}_\ell^{\rm BH}  
 =\mathcal{M}^{\ell+1}\left({\rm i } \chi\right)^\ell\,, \label{nohair}
\end{equation}
where $\mathcal{M}_\ell$ ($\mathcal{S}_\ell$) are the Geroch-Hansen mass (current) 
multipole moments~\cite{Geroch:1970cd,Hansen:1974zz}, the suffix ``BH'' refers to the BH
metric, $\mathcal{M}={\cal M}_0$ is the mass, $\chi\equiv{\mathcal{J} }/{\mathcal{M}^2}$ the dimensionless spin, and 
$\mathcal{J}=\mathcal{S}_1$ the angular momentum (we use natural units throughout). Equation~\eqref{nohair} implies 
that all Kerr moments with $\ell\geq2$ can be written only in terms of the mass $\mathcal{M}$ and angular momentum 
$\mathcal{J}$ of the spacetime.
Introducing the dimensionless quantities $\overline{{\cal M}}_\ell \equiv{\cal 
M}_\ell/{\cal M}^{\ell{+}1}$ and $\overline {\cal S}_\ell \equiv{\cal S}_\ell/{\cal M}^{\ell{+}1}$, the nonvanishing moments are
\begin{equation}
 \overline{{\cal M}}_{2n}^{\rm BH}   = (-1)^n \chi^{2n} \quad, \quad
 \overline{{\cal S}}_{2n{+}1}^{\rm BH} = (-1)^n \chi^{2n{+}1} \label{momKerr}
\end{equation}
for $n=0,1,2,...$. The fact that ${\cal M}_\ell=0$ 
(${\cal S}_\ell=0$) when $\ell$ is odd (even) is a consequence of the equatorial symmetry of the Kerr metric.
Likewise, the fact that all multipoles with $\ell\geq2$ are proportional to (powers of) the spin --~as well as their 
specific spin dependence~-- is a peculiarity of the Kerr metric,  that is lost for other compact-object solutions in 
GR~\cite{Pani:2015tga,Raposo:2018xkf} and also for BH solutions in other gravitational theories.

Testing whether these properties hold for an astrophysical dark object provides an opportunity to perform multiple 
null-hypothesis tests of the Kerr metric --~for example by measuring independently three multipole moments such as the 
mass, spin, and mass quadrupole $\mathcal{M}_2$~-- serving as a genuine strong-gravity 
test of Einstein's gravity~\cite{Psaltis:2008bb,Gair:2012nm,Yunes:2013dva,Berti:2015itd,
Cardoso:2016ryw,Barack:2018yly,Cardoso:2019rvt}, along with other proposed observational tests of fuzzballs (see, 
e.g., Refs.~\cite{Hertog:2017vod,Guo:2017jmi}). In this context it is intriguing that current 
gravitational-wave~(GW) observations (especially the recent GW190814~\cite{Abbott:2020khf} and 
GW190521~\cite{Abbott:2020tfl,Abbott:2020mjq}) do not exclude the 
existence of exotic compact objects other than BHs and neutron stars. 

In GR, BHs have curvature singularities that are conjectured to be always covered by 
event horizons~\cite{Penrose:1969pc,Wald:1997wa,Penrose_CCC}.
At the quantum level, BHs behave as thermodynamical systems with the area of the event horizon and its surface gravity 
playing the role of the entropy and 	temperature, respectively~\cite{Bekenstein,Hawking:1976de}. In fact a BH can 
evaporate emitting Hawking radiation~\cite{Hawking:1974sw}. This gives rise to a number of 
paradoxes that can be addressed in a consistent quantum theory of gravity such 
as string theory~\cite{Mathur:2009hf}. 

For special classes of extremal (charged BPS) BHs~\cite{Strominger:1996sh, Horowitz:1996ay, Maldacena:1997de} one can 
precisely count the microstates that account for the BH entropy. In some cases, one can even identify smooth horizonless 
geometries with the same mass, charges, and angular momentum as the corresponding BH. These geometries represent some of 
the microstates in the low-energy (super)gravity description. The existence of a nontrivial structure at the putative 
horizon scale is the essence of the fuzzball proposal~\cite{Lunin:2001jy, Lunin:2002qf, Mathur:2005zp, Mathur:2008nj}.
In the latter, many properties of BHs in GR emerge from 
an averaging procedure over a large number of microstates, or as a `collective behavior' of 
fuzzballs~\cite{Bianchi:2017sds, Bianchi:2018kzy, Bena:2018mpb, Bena:2019azk, Bianchi:2020des}. So far it has been hard 
to find a statistically significant fraction of microstate geometries both for five-dimensional (3-charge) and for 
four-dimensional (4-charge) BPS BHs. Yet, several classes of solutions based on a multicenter ansatz 
\cite{Bena:2015bea, Bena:2016agb, Bena:2016ypk, Bena:2017xbt, Bianchi:2017bxl, Bena:2017upb} have been found and their 
string theory origin uncovered~\cite{Giusto:2009qq, Giusto:2011fy, Bianchi:2016bgx}. 

Although in viable astrophysical scenarios BHs are expected to be neutral, charged BPS BHs are a useful toy model
to explore the properties of their microstates.
Extending the fuzzball proposal to neutral, non-BPS, BHs in four dimensions and finding predictions that can be 
observationally tested so as to distinguish this from other proposals and from the standard BH picture in 
GR~\cite{Cardoso:2019rvt} remain an open challenge. 

In this letter and in a companion paper~\cite{companion}, we investigate the differences in the multipolar structure 
between BHs and fuzzballs. As we shall argue, already at the level of the 
quadrupole moments the nonaxisymmetric geometry of generic microstates in the four-dimensional fuzzball model leads to 
a much richer phenomenology and to potentially detectable deviations from GR.

\noindent{{\bf{\em Setup.}}}
Our method is based on Thorne's seminal work on the multipole moments of a stationary isolated 
object~\cite{Thorne:1980ru}.
The idea is to choose a suitable coordinate system --~so called asymptotically Cartesian mass centered~(ACMC)~-- 
whereby the mass and current multipole moments can be directly extracted from a multipolar expansion of the metric 
components. In an ACMC system, the metric of a stationary asymptotically flat object can be written as~\cite{companion}
\be
\label{eq:ACMC}
ds^2= -(1-c_{00})dt^2 +c_{0i}\, dt \, dx_i +(1+c_{00})\, dx_i^2 +\ldots
\ee
with $x_i=\{x,y,z\}$, and $c_{00}$ and $c_{0i}$ admitting a 
spherical-harmonic expansion\footnote{It can be shown that the radial ($Y^{R}_{i,\ell m}$) and electric 
($Y^{E}_{i,\ell m}$) vector spherical harmonics only appear in subleading terms and do not affect the multipole 
moments~\cite{Thorne:1980ru}.} of the form~\cite{Thorne:1980ru} 
\begin{align}\label{cab_cart_coord}
c_{00} &{\,=\,} 2\sum_{\ell=0}^\infty \sum_{m=-\ell}^\ell \frac{1}{r^{1+\ell}}\sqrt{\frac{4\pi}{2\ell{\,+\,} 1}}  
\left( {\cal M}_{\ell m} Y_{\ell m} {\,+\,} \ell'{\,<\,}\ell \right) 
\\
c_{0i} &{\,=\,} 2 \sum_{\ell=1}^\infty\sum_{m=-\ell}^{\ell}\frac{1}{r^{1+\ell}}\sqrt{\frac{4\pi (\ell{\,+\,} 1)}{ \ell(2\ell{\,+\,} 1)}}   
\left( {\cal S}_{\ell m}Y^{B}_{i,\ell m}{\,+\,} \ell'{\,<\,}\ell \right) \nonumber
\end{align}
in terms of the scalar ($Y_{lm}$) and axial vector ($Y^{B}_{i,\ell m}$) spherical 
harmonics. 
The expansion coefficients ${\cal M}_{\ell m}$ and ${\cal S}_{\ell m}$ are the mass and current multipole 
moments of the spacetime, respectively. They can be conveniently packed into a single complex harmonic function
\begin{align} \label{HH}
H&= \sum_{\ell=0}^\infty \sum_{m=-\ell}^\ell \frac{1}{r^{1+\ell}}\sqrt{\frac{4\pi}{2\ell+1}}  
\left( {\cal M}_{\ell m} +{\rm i }  {\cal S}_{\ell m}  \right) Y_{\ell m}\,.
\end{align} 
In the case of the Kerr metric, $H$ is simply given by
\begin{align} \label{Hkerr}
  H_{\rm Kerr}=   \frac{ \cal M }{\sqrt{ x_1^2+x_2^2+\left(x_3-{\rm i} {\cal J\over \cal M}\right)^2 } }
\end{align} 
with two centers at positions $z=\pm {\cal J}/{\cal M}$ along the $z$-axis. 
The harmonic expansion of Eq.~\eqref{Hkerr} does not contain $m \neq 0$ terms, so that 
for each $\ell$ the moment tensors reduce\footnote{The normalization of Thorne's multipoles can be chosen in order 
to correspond to the Geroch-Hansen ones~\cite{Geroch:1970cd,Hansen:1974zz} used in 
Eq.~\eqref{nohair} in the axisymmetric case~\cite{Gursel}.} to the scalars ${\cal M}_{\ell}\equiv {\cal M}_{\ell0}$ and 
${\cal 
S}_{\ell}\equiv {\cal S}_{\ell0}$. The same holds for more general axisymmetric metrics.

\begin{table*}[t]
\begin{tabular}{c c c c c c c c c c c}
 \hline 
 \hline
Solution & $(\kappa_1, \kappa_2, \kappa_3,\kappa_4)$ & $\overline{\cal S}_{10} $  & $\overline{\cal M}_{20}$ 
&$\overline{\cal M}_{21}$  & $\overline{\cal M}_{22}$ & $\overline{\cal S}_{20}$ &$\overline{\cal S}_{21}$  & 
$\overline{\cal S}_{22}$ \\ 
 \hline 
\textbf{A} & (1,0,$k$,$k$)&0 & ${8\over 27 k^4}$ & $0$ & 0 & $0$ & 0 & $0$ \\
\textbf{B} &(1,0,1,$k$) & $ {L\over k}$  &  ${L^2\over k^2}$  & $0$ &  ${3 \sqrt{3} L^2 \over 
2\sqrt{2} k^3} $ 
 & $0$ & $-\frac{3 L^2}{\sqrt{2} k^3}$ & $0$\\
\textbf{C} &(3,0,$k$,2$k$)  &  ${4\sqrt{3}\, L\over 11^2\, k^3}$  & ${ 144\, L^2\over 11^4\, k^4}  $ & ${ 
72\,\sqrt{2}  L^2\over 11^4\, k^4} $ 
 & ${ 72\, \sqrt{6}\, L^2\over 11^4\, k^4}  $ 
 & $-\frac{164 L^2}{11^4 k^5}$ & $-\frac{48 \sqrt{2} L^2}{11^4 k^3}$ & $\frac{2 \sqrt{6} L^2}{11^3 k^3}$ 
\\
 \hline
 Kerr-Newman & & $\chi$ & $-\chi^2$ & $0$ &  $0$ &   $0$ &  $0$ &   $0$\\
 \hline 
 \hline
 \end{tabular}
 \caption{The first dimensionless multipole moments of some representative 3-center microstate 
geometries in the $k^2\gg L$. Moments with $m<0$ follow from $\overline{\cal M}_{\ell, -m}=(-1)^m \overline{\cal 
M}_{\ell, m}^{\,*}$.}
 \label{table:multipoles}
\end{table*}

Here we consider fuzzball solutions of gravity in four dimensions minimally coupled to four Maxwell fields and three 
complex scalars. A general class of extremal solutions of the Einstein-Maxwell system is described by a metric of the 
form~\cite{Bena:2007kg,Gibbons:2013tqa,Bates:2003vx} 
\begin{align} \label{4dsolution}
ds^2&=-e^{2U}\pt{dt+w}^2+e^{-2U}\sum_{i=1}^3 dx^2_i, 
\end{align}  
with 
\bea
e^{-4U} &{\,=\,}&  L_1\, L_2 \, L_3 \, V{\,-\,} K^1\, K^2\, K^3\, M {\,+\,}\frac12 \sum_{I>J}^3   K^I K^J  L_I L_J  \nn\\
&-& {M V\over 2}  \sum_{I=1}^3 K^I L_I  {\,-\,}\frac14 M^2 V^2
-\frac14 \sum_{I=1}^3 (K^IL_I)^2 \,, \nn\\
 *_{3}dw &=& \half\pt{VdM-MdV+K^IdL_I-L_IdK^I} \,,
\eea
where $*_3$ is the Hodge dual in 3-dimensional flat space, $ \{ V, L_I, K^I, M \}$ are eight harmonic functions 
associated with the four electric and four magnetic charges, and $I,J=1,2,3$.

Fuzzball solutions are obtained by distributing the charges of the eight harmonic functions among $N$ centers in such a 
way that the geometry near each center lifts to a regular five-dimensional geometry.  
More explicitly, we take
\bea
 V &=& v_0 + \sum_{a=1}^N  {v_{a}\over r_a }  \quad  ,  \quad  M = m_0 + \sum_{a=1}^N  {m_{a}\over r_a } \nn\\
  K^I &=& k^I_{0}+\sum_{a=1}^N  {k^I_{a}\over r_a } \quad  ,  \quad      L_I = \ell_{I,0}+ \sum_{a=1}^N   { 
{\ell}_{I,a} \over  r_a }    \label{ansatz0}
 \eea
with $r_a =| {\bf x}-{\bf x}_a|$ the distance from the $a$-th center.  

\noindent{{\bf{\em Results.}}}
%
Comparing the metric~\eqref{4dsolution} with the definition of an ACMC metric~\eqref{eq:ACMC}, one can 
extract the multipole moments of the fuzzball solution (details are given in Ref.~\cite{companion}). 
The fuzzball multipole moments are encoded in the multipole harmonic function 
\begin{equation}
H={1\over 4} \sum_{a=1}^N \left[ V+{\rm i} M +\sum_{I=1}^3 (L_I-{\rm i}\, K^I)   \right]\,.
\end{equation}
This complex harmonic function is a generalization of the Kerr case
[Eq.~\eqref{Hkerr}], the latter can be interpreted as a two-center solution, with the Schwarzschild case corresponding 
to a single center. The above expression is instead valid for generic $N$-center solutions, regardless of the presence of electromagnetic and scalar fields.

Expanding the harmonic function $H$ yields the multipole moments
\begin{equation}\label{msresults}
\begin{aligned}
{\cal M}_{\ell m} & = \frac{1}{4}\sum_{a=1}^N\left(v_a+\sum_I\ell_{I,a}\right)R_{\ell m}^a, \qquad
\ell \geq 0   \quad  \\
{\cal S}_{\ell m}&= \frac{1}{4}\sum_{a=1}^N\left(m_a-\sum_I k_a^I\right)R_{\ell m}^a, \qquad\ell \geq 1
\end{aligned}
\end{equation}
with ${\cal M}_{00}={\cal M}$ and 
\begin{equation}
R_{\ell m}^a = |\textbf{x}_a|^\ell \sqrt{\frac{4\pi}{2\ell+1}} Y_{\ell m}^*(\theta_a,\phi_a)\,.
\end{equation}
As in the case of axisymmetric geometries, we define dimensionless moments
\begin{equation}
\overline{\mathcal{M}}_{\ell m} = \frac{\mathcal{M}_{\ell m}}{\mathcal{M}^{\ell+1}}
\quad
,
\quad
\overline{\mathcal{S}}_{\ell m} = \frac{\mathcal{S}_{\ell m}}{\mathcal{M}^{\ell+1}}\,.
\end{equation} 
We center the coordinate system in the center-of-mass and orient the $z$-axis along the angular momentum, so  that
\begin{equation}
 \begin{aligned}
 \frac{1 }{4} \sum_{a=1}^N  \left(v_a+\sum_I\ell_{I,a}\right) {\bf x}_a  &=  0  \\
  \frac{1 }{4}\sum_{a=1}^N  \left(m_a-\sum_I k_a^I\right) {\bf x}_a  &=  {\cal J}{\bf e}_z
\end{aligned}\quad\,,
\end{equation}
with ${\bf e}_z$ the unit vector along $z$. With this choice ${\cal M}_{1m}=0$, ${\cal S}_{1,\pm 1}=0$, 
and ${\cal S}_{10}={\cal J}$. 
  
Equations~\eqref{msresults} are one of our main results, as they allow us to compute the multipole moments of  any 
multicenter microstate geometry. In fact, our method can be straightforwardly applied to \emph{any} metric in ACMC 
form. In the following we will focus on some specific cases.

\noindent{{\bf{\em Examples.}}}
%
The simplest horizonless geometries arise from three-center solutions. We consider 
fuzzballs that asymptote to BHs carrying three electric ($Q_I$) and one magnetic ($P_0$) charge, obtained from 
orthogonal branes, so we require that $K^I$ and $M$ vanish at order $1/r$. Up to a reordering of the centers, the 
general solution can be written in the form~\cite{Bianchi:2017bxl}
\bea
V &=& 1+\sum_{a=1}^3 {1\over r_a}   \qquad   M=  \kappa_1 \,\kappa_2 \,\kappa_3\, \kappa_4  \left( { 1\over r_1}-{1\over r_2}\right) 
\\
 L_1 &=&1+  \kappa_4 \left( {\kappa_3 \over r_1}{\,-\,}{\kappa_2\over r_2} \right)  \qquad  L_2 =1+  \kappa_1 \kappa_4 \left( {\kappa_3\over r_2} {\,-\,}{\kappa_2\over r_1} \right) \nn \\
 L_3 &=&1+  \kappa_1 \left( {\kappa_2\kappa_3 \over r_1}+{\kappa_2 \kappa_3\over r_2} +{(\kappa_2+\kappa_3)^2 \over r_3} \right) \nn\\
K_1 &=&  \kappa_1 \left( -{\kappa_2 \over r_1}-{\kappa_3\over r_2}+{\kappa_2+\kappa_3\over r_3} \right) \nn\\
  K_2 &=&  {\kappa_3 \over r_1}+{\kappa_2 \over r_2}-{\kappa_2+\kappa_3 \over r_3}  \qquad 
  K_3 =  \kappa_4 \left({1\over r_2}-{1\over r_1} \right)  \nn
 \eea
 with $\kappa_\alpha$ some arbitrary integers. 

Regular solutions describe microstates of a (nonrotating) BPS BH with mass 
\begin{equation}
\mathcal{M} = \frac{1}{4}\left(Q_1+Q_2+Q_3+P_0\right)
\end{equation}
and charges 
\bea
Q_1&=& \kappa_4(\kappa_3-\kappa_2)\,,\qquad Q_2= \kappa_1\kappa_4(\kappa_3-\kappa_2)\,,\nn\\
Q_3 &=& \kappa_1(\kappa_2^2+4\kappa_2\kappa_3+\kappa_3^2)\,,\qquad P_0=3\,.
\eea

Besides the integer parameters $\kappa_\alpha$, the solution depends on some continuous 
parameters, namely, the distances between the centers $r_{ab}{=}| {\bf x}_a{-}{\bf x}_b|$. These are constrained by 
the so-called `bubble equations' \cite{Bena:2007kg}, ensuring regularity of the five-dimensional lift and absence of 
closed time-like curves. In the 3-center case one has %
\bea
\small
r_{12}&=& \frac{2\kappa_1\kappa_4(\kappa_2{-}\kappa_3)^2r_{23}}{\kappa_1\kappa_4(2\kappa_2^2{+}5\kappa_2\kappa_3{+}2\kappa_3^2){+}[\kappa_2{+}\kappa_4{-}\kappa_1\kappa_3(1{-}\kappa_2\kappa_4)]r_{23}}\nn\\
r_{13}&=& \frac{\kappa_1\kappa_4(2\kappa_2{+}\kappa_3)(\kappa_2{+}2\kappa_3)r_{23}}{\kappa_1\kappa_4(2\kappa_2^2{+}5\kappa_2\kappa_3{+}2\kappa_3^2){-}(\kappa_1{-}1)(\kappa_2{+}\kappa_3)r_{23}}
\eea
which allow one to express $r_{12}$ and $r_{13}$ in terms of $r_{23}=L$, the surviving continuous parameter (`modulus') 
labeling the microstate.
Asymptotically the solution coincides with the Kerr-Newman metric~\cite{KN}, whose multipolar structure is the same~\cite{Sotiriou:2004ud} as in the Kerr case [see Eq.~\eqref{nohair}].

A summary of the first multipole moments for some representative cases is shown in Table~\ref{table:multipoles}. 
The general expressions for the multipole moments are cumbersome so we present them in the limit of 
large mass ($\kappa_\alpha\gg1$), which is also the most interesting one from a phenomenological point of view, since 
it corresponds to objects with mass arbitrarily larger than the Planck mass. 
We consider three representative arrangements of the three centers:
\begin{itemize}
\item{{\bf A}:\,\emph{Equilateral triangle}.  $(\kappa_1,\kappa_2,\kappa_3,\kappa_4)$$\,=\,$$(1,0,k,k)$. 
 These microstate geometries fall into the class of ``scaling solutions'' characterized by zero angular 
momentum, $\mathcal{J} = 0$, equal charges $\vec{Q}=(k^2,k^2,k^2)$, and mass ${\cal M}={3\over 4}(1+k^2)$.
Thanks to ${\bf Z}_3$ symmetry around $z$, the nontrivial mass multipole moments read 
\begin{equation}
{\cal M}_{2p+3n,3n} =  {\cal M} (-L)^{2p+3n} \frac{\sqrt{(2p+6n)!(2p)!}}{2^{2p+3n} (p+3n)! n!}
\end{equation}
where $L=r_{12}=r_{23}=r_{31}$. 
Thus, at variance with the Kerr case, the mass quadrupole moments are not spin induced: 
they can be nonzero even if the spin ${\cal J}$ vanishes. Furthermore, for $\ell\geq3$ they also have 
$m\neq0$ components of the mass moments. The large $k$ limit of all quadrupole moments are displayed 
in Table~\ref{table:multipoles}.
}
\item{{\bf B}:\,\emph{Isosceles triangle}.  $(\kappa_1,\kappa_2,\kappa_3,\kappa_4)=(1,0,1,k)$. 
These microstate geometries possess non vanishing angular momentum, $\mathcal{J} = \frac{(k-1) k L}{2[k (L+2)-L]}$, 
charges $\vec {Q} = (k,k,1)$, and mass $\mathcal{M} = \frac{2+k}{2}$. In this case $L=r_{23}=r_{31}>r_{12}$. For 
$k\to\infty$ and $L\ll1$ (see Table~\ref{table:multipoles}), the multiple moments coincide with those of the Kerr 
metric modulo the factors $(-1)^n$ in Eq.~\eqref{momKerr}. In particular, while 
the Kerr metric is oblate (${\cal M}_2<0$), these solutions are prolate (${\cal M}_2>0$). However, for finite values of 
$k$ the solution also displays quadrupole moments that break axial symmetry, e.g. ${\cal M}_{22}$ and ${\cal S}_{21}$.
} 
\item{{\bf C}:\,\emph{Scalene triangle}.  $(\kappa_1,\kappa_2,\kappa_3,\kappa_4)=(3,0,k,2k)$. These microstate 
geometries possess a non vanishing angular momentum $\mathcal{J}$ which is a complicated function of $k$ and $L$, with 
$L=r_{23}< r_{12}<r_{31}$, charges $\vec
{Q} = (2k^2,6k^2,3k^2)$, and mass $\mathcal{M} = \frac{3+11k^2}{4}$. For large $k$ one finds $\mathcal{J}\sim 
\frac{\sqrt{3}k L}{4}$.
Triangle inequalities require $\nu\equiv{L\over 12 
k^2}<1{-}{1\over \sqrt{2}}$. The multipole moments for large $k$ are displayed in Table~\ref{table:multipoles}. In this 
case both the axisymmetry and the equatorial symmetry of the Kerr metric are broken, as shown by the fact that the 
multipole moments ${\cal M}_{\ell m}$ and ${\cal S}_{\ell m}$ are generically nonzero.
} 
\end{itemize}

\begin{figure}[t]
\centering
 \includegraphics[scale=0.25]{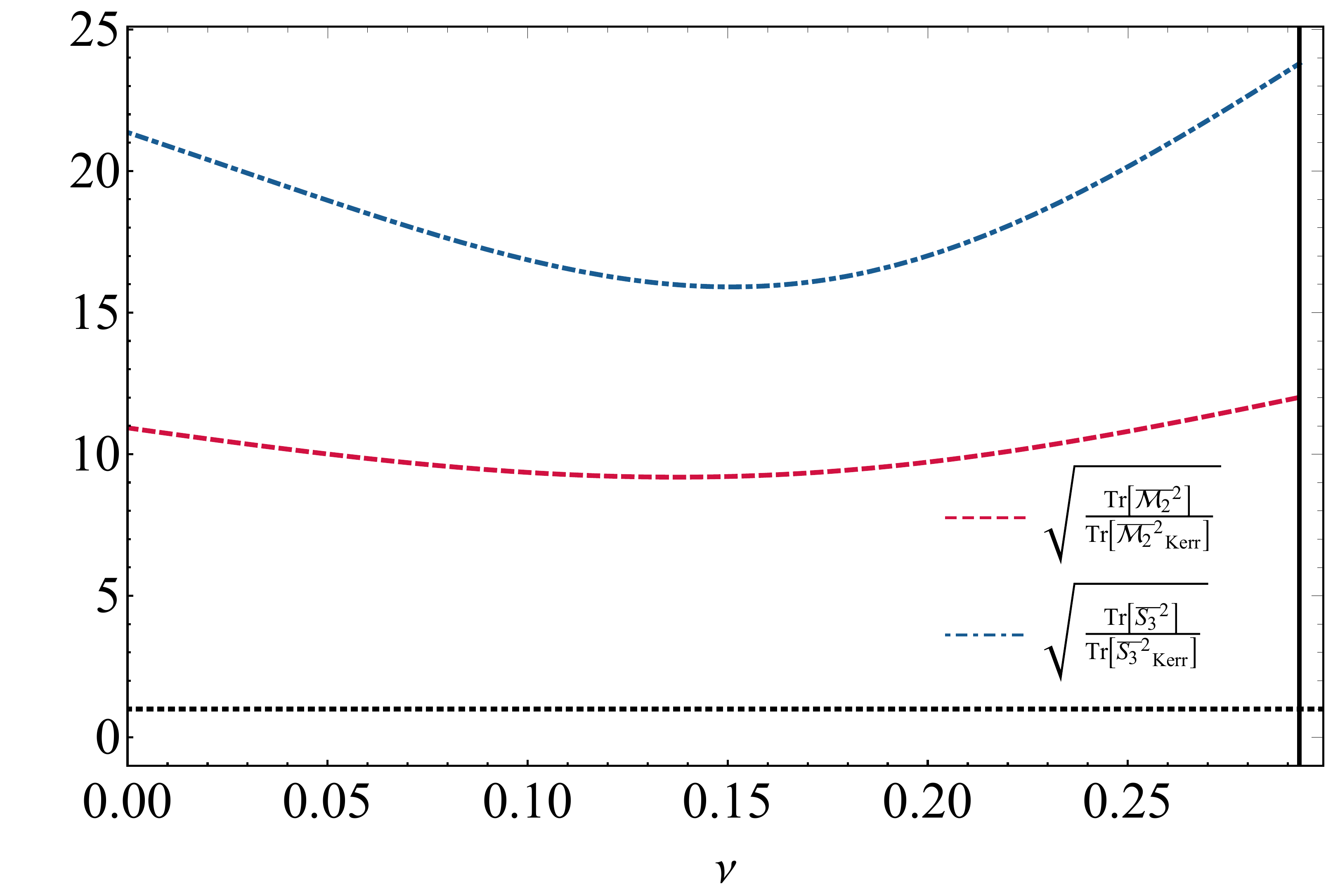} 
\caption{Ratios between the quadratic invariants for the first multipole moments of a fuzzball (solution 
$\bf C$) and a Kerr BH with the same angular momentum, as a function of $\nu = L/(12k^2)$ with $k=1$. The vertical 
solid line corresponds to the upper bound $\nu_{\rm max}=1-1/\sqrt{2}$. The horizontal dotted black line refers to the 
fuzzball and Kerr moments being identical. The fuzzball moments are larger than the corresponding Kerr ones, which is a 
typical property~\cite{companion}.
}
\label{plotSM}
\end{figure}

It is interesting to observe that the mass and current multipole moments of these microstate geometries are typically 
larger than those of a Kerr-Newman BH with same mass and angular momentum. A representative 
example of this property is shown in Fig.~\ref{plotSM}, where we display some ratios between multipole moments of 
microstate geometries of type ${\bf C}$ and those of a Kerr BH.  
We focus on the quadratic invariants
\begin{equation}
 {\rm tr} {\cal M}_\ell^2 =  \sum_{m=-\ell}^\ell | {\cal M}_{\ell m} |^2\,,\qquad {\rm tr} S_\ell^2 = 
\sum_{m=-\ell}^\ell |S_{\ell m} |^2\,.
\end{equation}
We have explored numerically a large region of the whole $(\kappa_\alpha,L)$ parameter space and found that quadratic 
invariants for the microstate geometries are typically bigger than those of Kerr BHs for any $\ell$~\cite{companion}. 
It would be 
interesting to find a general proof of this property, which is analogous to the fact that the Lyapunov exponent of 
unstable null geodesics near the photon sphere is maximum for the BH solution~\cite{Bianchi:2020des}. In other words, 
both for the multipole moments and for the Lyapunov exponent, the BH solution appears to be an extremum point in the 
space of the solutions.

\noindent{{\bf{\em Phenomenological implications.}}}
%
The above examples are representatives of some general features of this large family of solutions. In particular, the 
$\ell\geq2$ multipole moments of fuzzball geometries are not necessarily spin induced as in the Kerr case, they can 
break axial and equatorial symmetries, and are larger than in the Kerr case.
The peculiar multipolar structure and the striking deviation from the Kerr multipoles provides a portal to constrain 
fuzzball models with current and future observations, with both electromagnetic and GW probes~\cite{Cardoso:2019rvt}. 

By analyzing the accretion flow near the supermassive BH in M87, the Event Horizon 
Telescope placed a mild bound on its dimensionless (axisymmetric) quadrupole moment, $|\overline{\cal 
M}_2-\overline{\cal  M}_2^{\rm BH}|\lesssim 
4$~\cite{Akiyama:2019cqa}. Furthermore, in a coalescence the quadrupole moment of the binary components 
affect the GW signal through a next-to-next to leading post-Newtonian 
correction~\cite{Blanchet:2006zz,Krishnendu:2017shb}. 
Constraints on parametrized post-Newtonian deviations using the events from the first LIGO-Virgo 
Catalog~\cite{LIGOScientific:2018mvr,LIGOScientific:2019fpa} can be mapped into a constraint $|\overline {\cal 
M}_2-\overline{\cal  M}_2^{\rm BH}|\lesssim 1$, in particular using the 
events GW151226 and GW170608~\cite{DiPasquaInPrep}. 
Comparing with deviations found in the microstate solutions, current bounds are not particularly 
stringent.

While current GW constraints will become slightly more stringent in the next years as the sensitivity of the 
ground-based detectors improve~\cite{Krishnendu:2018nqa}, much tighter bounds will come from extreme mass-ratio 
inspirals~(EMRIs), one of the main targets of the future space mission LISA~\cite{Audley:2017drz}.  Although EMRI data 
analysis is challenging~\cite{Babak:2017tow,Chua:2018yng,Chua:2019wgs,LISADataChallenge} the potential reward is unique: 
a detection of 
these systems can be used to measure the ($m=0$, mass) quadrupole moment $\overline{{\cal M}}_2$ of the central 
supermassive object with an accuracy of one part in $10^4$~\cite{Barack:2006pq,Babak:2017tow}, offering unprecedented 
tests of exotic compact objects~\cite{Glampedakis:2005cf,Raposo:2018xkf,Destounis:2020kss}. 

While our results suggest that very strong constraints on fuzzball geometries can be set with EMRIs, a precise analysis 
requires a class of neutral, nonextremal solutions, which would further imply the absence of extra emission 
channels (e.g. dipolar radiation). For astrophysically viable objects, we expect that the multipolar structure is the 
only discriminant with respect to the Kerr BH case, which can be explored with the methods presented here.

In addition to having a different quadrupole moment, microstate geometries are much less symmetric than the Kerr 
metric, which implies the existence of multipole moments that are identically zero in the Kerr case (see also 
Refs.~\cite{Ryan:1995wh,Raposo:2018xkf}). Investigating how multipole moments that break equatorial 
symmetry or axisymmetry (e.g., ${\cal S}_{2m}$ and ${\cal M}_{2m}$ with $m\neq0$) affect the GW waveform and their 
phenomenological consequences is an important topic that is left for a follow-up work.

Finally, a broad statistical analysis shows that certain invariant combinations of the $\ell\geq2$ multipole moments of 
three-center microstate geometries are larger than those of the corresponding Kerr BH in a wide region of the 
four-dimensional parameter space, and are always larger than their corresponding value in the $L\to0$ 
limit~\cite{companion}.
If confirmed, this result would imply that any future measurement of the invariant combinations of the multipole moments 
smaller than the BH ones can potentially rule out this family of solutions to be
typical microstates of the corresponding BH, with important consequences for the fuzzball scenario.

\noindent{{\bf{\em Note added.}}}
While this work was in preparation, a related work by Iosif Bena and Daniel R.~Mayerson appeared~\cite{Bena:2020see} 
(see also the more recent companion~\cite{Bena:2020uup}). 
The idea and aims of that paper are similar to ours. Ref.~\cite{Bena:2020see} focuses on axisymmetric geometries in the 
BH limit, whereas our results are valid beyond axial symmetry in regions where the microstate geometries can 
significantly deviate from the BH metric.

\noindent{{\bf{\em Acknowledgments.}}}
D.C. was supported by FWF Austrian Science Fund via the SAP P30531-N27.
P.P. acknowledges financial support provided under the European Union's H2020 ERC, Starting 
Grant agreement no.~DarkGRA--757480, and under the MIUR PRIN and FARE programmes (GW-NEXT, CUP:~B84I20000100001), and 
support from the Amaldi Research Center funded by the MIUR program ``Dipartimento di Eccellenza'' (CUP: 
B81I18001170001).
%

\bibliographystyle{utphys}
\bibliography{Ref}

\end{document}